\documentclass[a4paper]{jpconf}
\usepackage{graphicx}

\begin{document}
\title{The ArDM project: a Dark Matter Direct Detection Experiment based on Liquid Argon}

\author{Lilian Kaufmann, Andr\'e Rubbia}

\address{Institute for Particle Physics, ETH Zurich}

\ead{lilian.kaufmann@cern.ch, andre.rubbia@cern.ch}

\begin{abstract}
The Dark Matter part of the universe presumably consists of WIMPs (Weakly Interacting Massive Particles). The ArDM project aims at measuring signals induced by WIMPs in a liquid argon detector. A 1-ton prototype is currently developed with the goal of demonstrating the feasibility of such a direct detection experiment with large target mass. The technical design of the detector aims at independently measuring the scintillation light and the ionization charge originating from an interaction of a WIMP with an argon nucleus. The principle of the experiment and the conceptual design of the detector are described. 
\end{abstract}

\section{Introduction}

Astronomical observations indicate the presence of a large amount of invisible matter in our universe. Presumably, over 80\% of the total mass content of the universe consists of so-called Dark Matter. This kind of matter interacts only weakly with ordinary matter, and manifests itself mainly via gravitational influence on visible structures like galaxies or galaxy clusters. Common models for the distribution of Dark Matter describe it as consisting of a sea of heavy, weakly interacting, non-baryonic and massive particles, or WIMPs (Weakly Interacting Massive Particles), forming a cold thermal relic gas. \\ 
A WIMP could interact with a target detector material via elastic scattering, imparting nuclear recoils in the energy range between 10 keV and 100 keV. Measuring these recoils is an approach for directly detecting Dark Matter. Due to the weak coupling and the low imparted energy, WIMP signals are however rare and elusive events. The total interaction rate depends on underlying models, e.\,g.~supersymmetric models, whose parameters are numerically poorly known, and is therefore not yet predictable. \\ 
One category of target materials are the liquid noble elements like argon or xenon, providing advantageous features for detection like high density and a low ionization potential, resulting in a high ioniziation and scintillation yield. Charge from ionization and light from scintillation provide two different signals caused by interactions of particles with the detector material. By considering the typical relative charge and light yields induced by different types of interactions, the particles causing events in the detector can be identified. Additionally, the time dependence of the scintillation light can be used to further discriminate between WIMP signals and background signals. Nevertheless, it is important to reduce background radiation to a very low level. WIMP Direct Detection experiments therefore need to be located underground, and great care has to be taken when choosing the detector materials. \\
In 2004, the Argon Dark Matter Experiment (ArDM) was initiated. The goal of this project is the development of a 1-ton argon detector with independent ionization and scintillation readout, demonstrating the feasibility of a ton-scale liquefied noble gas Dark Matter experiment. 

\section{Design of the ArDM detector}

\begin{figure*}[!ht]
\begin{center}
\includegraphics*[width=11cm,angle=-90]{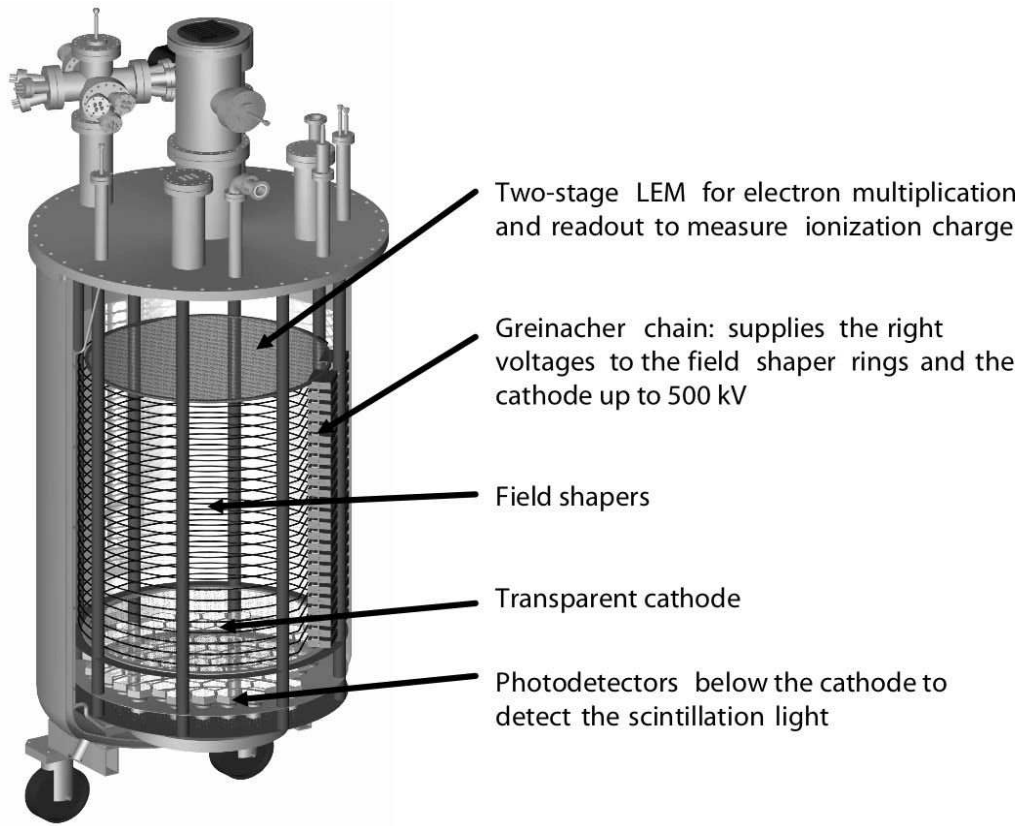}
\caption{Setup of the detector.}
\end{center}
\label{fig}
\end{figure*}

\noindent
Fig.~\ref{fig} illustrates the conceptual design of the ArDM 1-ton prototype \cite{Talk,Proc}. Its three technical keypoints are the charge readout on the top, the drift field and the light readout on the bottom of the detector. 
After an event in the fiducial volume has produced scintillation light and free ionization electrons, the electrons are pulled away from the interaction point and drifted upwards. The drift field is provided by a Greinacher high-voltage circuit, designed to reach up to 4~kV/cm. The maximal drift length is approximately 120~cm. When the electrons reach the level of the liquid argon surface, they are extracted into a gaseous argon volume on the top of the detector where the readout of the ionization charge takes place. \\
In order to produce a detectable signal, the electrons need to be multiplied, even in a sensitive charge preamplifier (typical equivalent noise $\simeq$ 1000 electrons). In argon, such a charge amplification is achieved in the gas phase by using two LEM (Large Electron Multiplier) plates on top of each other for charge amplification and readout. The readout is segmented, defining the location of the event in space together with the prompt light signal. \\ 
For the readout of the scintillation light, an array of 14 photomultiplier tubes is used, which is located at the bottom of the detector. The primary VUV scintillation light of argon (wavelength 128 nm) needs to be shifted to visible light in order to match the sensitivity range of the photomultiplier tubes. This is achieved by a TPB (Tetra-Phenyl-Butadiene) compound on a reflecting polymer, acting as a wavelength shifter. \\
The combination of charge and light readout should provide a powerful background rejection possibility. Since nuclear recoils due to WIMP or neutron interactions have a much higher characteristic light over charge ratio than electron and gamma interactions, these types of events are in principle well distinguishable if the light over charge ratio can be measured precisely. This is a mandatory condition for the rejection of predominant background signals like the internal $^{39}$Ar signal. The $^{39}$Ar isotope is present in natural argon liquefied from the atmosphere \cite{Loosli}, and produces a background rate due to beta decay of approximately 1 kHz in one ton of liquid argon. The alternative possibility of using $^{39}$Ar-depleted argon extracted from underground natural gas is currently also studied. \\
The comparison of fast and slow light components can also be used to further discriminate between nuclear recoils and background events. Due to different ionization densities, the population of argon excimer states is different for nuclear recoil and electron/gamma events, leading to characteristic contributions to the fast and slow scintillation light components. \\
The choice of natural argon instead of xenon for the initial ton-scale target can be motivated by three arguments: 

\begin{itemize}
\item{The event rate in argon is less sensitive to the energy threshold than in xenon, due to form factor effects.} 

\item{Argon is cheaper than xenon, and sizeable experience in the handling of massive liquid argon detectors has been acquired within the ICARUS program \cite{IC1,IC2}. A ton-scale argon detector is hence readily conceivable, safe and economically affordable.} 

\item{Recoil spectra are different in xenon and argon, providing a crosscheck if a potential WIMP-signal is measured.} 
\end{itemize}

\section{Outlook}

Over the last years, a lot of efforts have been undertaken to find out more about the nature of dark matter as well as the particle which it might consist of. It is believed that liquefied noble gas detectors provide one approach to identify the WIMP. The ArDM project aims at the application of liquid argon for the direct detection of nuclear recoils induced by WIMPs. The first goal of the project is the construction of a 1-ton prototype in order to demonstrate the feasibility of the experiment. The three keypoints for a successful operation are:

\begin{itemize}
\item{LEM-based charge readout in gaseous argon}
\item{Generation of a very high drift field}
\item{Efficient readout of the argon scintillation light} 
\end{itemize}

The 1-ton prototype is currently under construction at CERN. Its operation involves the design, acquisition and run of a cryogenic system, a liquid argon purification system, a high-voltage system, charge amplification and readout, and light readout. 
The first milestone to be achieved during the first months of 2007 is a proof of principle and stability studies, as well as optimization of the detector performance regarding $\gamma$-ray and beta electron background rejection versus nuclear recoils. In a second phase, the design of a shielding against neutrons from detector components, surrounding facilities and muon-induced neutrons will be addressed. \\ After the successful construction, testing and operation of the prototype, an underground operation is planned in order to minimize background radiation induced by cosmic rays. \\
With a recoil energy threshold of 30 keV, a WIMP-nucleon cross section of $10^{-42}$ cm$^2$ would yield approximately 100 events per day per ton. The sensitivity of the ArDM 1-ton prototype would therefore access the WIMP-nucleon cross section region of $10^{-6}$ pb. By improving the background rejection power and further limiting the background sources, a sensitivity of $10^{-8}$ pb would become reachable. Scaling linearly with mass, a 10-ton detector could reach a ten times smaller cross section. Due to the scalable technologies used and the low cost of argon, an enlargement of the volume is a realistic prospect. 


\section{Acknowledgements}

The help of all ArDM colleagues from ETH Z\"urich, Granada University, CIEMAT, Soltan Institute Warszawa, University of Sheffield and University of Z\"urich is greatly acknowledged. Informal contributions from P. Picchi (LNF) are also greatly recognized.

\end{document}